\documentclass[11pt]{article}
\usepackage[english]{babel}

\usepackage[a4paper,top=2cm,bottom=2cm,left=2.5cm,right=2.5cm,marginparwidth=1.75cm]{geometry}

\usepackage{setspace}
\usepackage{amsmath}
\usepackage{graphicx}
\usepackage[colorlinks=true, allcolors=blue]{hyperref}

\title{ESO Expanding Horizons White Paper: The 15-m Galactic Archaeology Spectroscopic Surveyor}

\author{
  Borja Anguiano$^{1}$, First Author$^{2}$, Second Author$^{3}$\\[0.5em]
  {\small
  $^{1}$Centro de Estudios de F\'isica del Cosmos de Arag\'on (CEFCA), Spain\\
  $^{2}$Affiliation 2 (ESO Member State)\\
  $^{3}$Affiliation 3 (ESO Member State)
  }
}

\date{\today}

\usepackage{etoolbox}

\makeatletter
\patchcmd{\abstract}{\section*{\abstractname}}{}{}{}
\makeatother

\begin{document}

\begin{titlepage}
    \centering

    {\LARGE\bfseries Why the Northern Hemisphere Needs a 30–40 m Telescope and the Science at Stake: \\ \vspace{.5cm} Galactic Archaeology from the Northern Sky\par}
    \vspace{1.5cm}

    {\large
    Borja Anguiano$^{1}$, David Valls-Gabaud$^{2}$, Guillaume F. Thomas$^{3,}$$^{4}$, David Mart\'inez Delgado$^{1}$, Alberto M. Mart\'inez-Garc\'ia$^{5}$, Andr\'es del Pino$^{5}$, Ivan Minchev$^{6}$, Patricia Sanchez-Blazquez$^{7}$, Carme Gallart$^{3,}$$^{4}$, Teresa Antoja$^{8}$\par}
    \vspace{0.5cm}

    {\small
    $^{1}$Centro de Estudios de F\'isica del Cosmos de Arag\'on (CEFCA), Plaza San Juan 1, 44001 Teruel, Spain.\\
    $^{2}$ LERMA, CNRS, Observatoire de Paris, 61 Avenue de l’Observatoire, Paris, 75014, France.\\
    $^{3}$ Instituto de Astrofísica de Canarias, 38205, La Laguna, Tenerife, Spain. \\ 
    $^{4}$ Universidad de La Laguna, Dpto. Astrofísica, 38206, La Laguna, Tenerife, Spain. \\
    $^{5}$ Instituto de Astrofísica de Andalucía, CSIC, Glorieta de la Astronomía, 18080, Granada, Spain. \\
    $^{6}$ Leibniz-Institut für Astrophysik Potsdam (AIP), An der Sternwarte 16, 14482, Potsdam, Germany \\
    $^{7}$ Departamento de Física de la Tierra y Astrofísica \& IPARCOS, Universidad Complutense de Madrid, 28040, Madrid, Spain \\
    $^{8}$ Institut de Ciences ` del Cosmos, Universitat de Barcelona, Barcelona 08028, Spain\par}

    \vfill

    {\large \today\par}
\end{titlepage}

\setcounter{page}{1}


\begin{abstract}
By the 2040s--50s, facilities such as \emph{Gaia}, WEAVE, 4MOST, Rubin, \emph{Euclid}, \emph{Roman}, and the ESO ELT will have transformed our global view of the Milky Way. Yet key questions will remain incompletely resolved: a detailed reconstruction of the Galaxy's assembly from its earliest building blocks, and robust tests of dark matter granularity using the fine structure of the stellar halo and outer disk---particularly in the Galactic anticenter. Addressing these questions requires high-resolution spectroscopy of faint main-sequence stars (typically 1--2 mag below the turnoff) and turnoff stars ($r \sim 21$--23) in low-surface-brightness structures: halo streams and shells, ultra-faint dwarf galaxies, the warped and flared outer disk, and anticenter substructures. We argue that addressing this science case requires a 30\,m-class telescope in the northern hemisphere, equipped with wide-field, highly multiplexed, high-resolution spectroscopic capabilities. Such a facility would enable (i) a Northern Halo Deep Survey of $\sim 10^{5}$--$10^{6}$ faint main-sequence and turnoff stars out to $\sim 150$--200\,kpc, (ii) chemodynamical mapping of dozens of streams to measure perturbations from dark matter subhalos, and (iii) tomographic studies of the anticenter and outer disk to disentangle perturbed disk material from accreted debris. A northern 30\,m telescope would provide the essential complement to ESO's southern ELT, enabling genuinely all-sky Milky Way archaeology and delivering stringent constraints on the small-scale structure of dark matter.
\end{abstract}

\section{Introduction and Motivation}

By the 2040s the Milky Way will be mapped in unprecedented depth and dimensionality. Astrometric and photometric surveys such as \emph{Gaia} [1], (potentially \emph{GaiaNIR} [2]), Rubin [3], \emph{Euclid} [4], and \emph{Roman} [5] will have mapped billions of stars and low-surface-brightness structures across the sky, while large spectroscopic efforts (e.g., LAMOST, WEAVE, 4MOST, SDSS-V) will have provided radial velocities and medium-resolution abundances for tens of millions of relatively bright targets. The ESO Extremely Large Telescope (ELT) [6] will deliver exquisite, but largely targeted, high-resolution spectroscopy and resolved imaging of selected regions of the Galaxy, particularly in the southern hemisphere. Together, these facilities will establish an impressive global picture of the Milky Way's structure and kinematics. However, some of the most fundamental questions about the Galaxy's \emph{detailed} assembly history and the \emph{granularity} of its dark matter halo will remain only partially addressed. In particular, current and planned facilities are not optimized to obtain high-resolution, high signal-to-noise spectra for faint main-sequence stars (typically 1--2 mag below the turnoff) and turnoff stars that dominate the outer halo, stellar streams, ultra-faint dwarf galaxies (UFDs), and outer disk ---best targetted from the northern sky in the direction of the Galactic anticenter.

\noindent\textbf{Central question:} \emph{How did the Milky Way assemble, in detail, from its earliest building blocks, and what does the fine structure of its stellar halo, including the surviving UFDs, and outer disk---particularly in the Galactic anticenter---reveal about the nature and granularity of dark matter?}

The key fossil record of the Milky Way’s formation is encoded in the chemodynamical properties of these stars. Their multi-element abundance patterns trace the conditions in their birth environments and allow us to chemically tag distinct accretion events [7]. The larger building blocks, such as Gaia--Enceladus [8], contain a large number of bright targets that are allowing a fair study with current or upcoming facilities. Their full phase-space distribution, including precise radial velocities, constrains the Galaxy’s gravitational potential and its small-scale structure, including perturbations from substructure as inferred from stream dynamics [9,10]. In addition to the accreted and UFD systems themselves, many of the most informative outer-disk and anticenter substructures can be connected to disk perturbations and warp/flare phenomena [11,12]. However, for smaller or low surface brightness systems (which census is expected to grow substantially in the coming years [13]), observing faint stars is crucial. A substantial step forward in these issues requires deep, highly multiplexed ground-based spectroscopy over degree-scale fields. We therefore argue that addressing this central question requires a 30\,m-class telescope equipped with wide-field, highly multiplexed, high-resolution spectroscopic capabilities. In the same way as the observation of the Galactic Center has favoured the construction of the ELT and other observing facilities in the South, placing such new facility in the Northern Hemisphere is warranted by the need to study in detail the Galactic anticenter, in addition to the streams and UFD equally observable from both Hemispheres. With MSE no longer moving forward, there is a timely opportunity to pursue a northern facility that can deliver wide-field, highly multiplexed, high-resolution spectroscopy for Milky Way archaeology at the required scale [14]. In practical terms, it would enable survey-scale acquisition of $R\sim 30{,}000$ spectra for faint main-sequence and turnoff stars over wide northern fields (halo, streams, and the anticenter). Without such a capability, progress would rely mainly on small, targeted samples that cannot reach the required statistics. At the same time, such a facility would be much better suited for targeted follow-up of particularly valuable tracers identified by these surveys, including extremely metal-poor stars and the most distant blue horizontal-branch stars. It would also ensure that the northern sky contributes on equal footing to the ELT era, enabling genuinely all-sky chemodynamical mapping and maximizing the scientific return of \emph{Gaia}, Rubin, \emph{Euclid}, and \emph{Roman}.

\section{Why a Northern 30 m-class Telescope is Essential}

The science case in Section 1 requires a combination of depth, spectral resolution, multiplexing, and sky coverage, best delivered by a Northern Hemisphere 30 m–class telescope with optimal access to the Galactic anticenter.

\paragraph{Faint main-sequence stars as a primary fossil record.}
The brightest giants have been and will remain indispensable tracers of the distant halo and stellar streams. However, a complete reconstruction of the Galaxy’s assembly history requires large samples of faint main-sequence stars (typically 1--2 mag below the turnoff) and turnoff stars, particularly in the outer halo and anticenter. These stars are far more numerous, preserve relatively unprocessed chemical abundance patterns, are less affected by internal mixing, and provide significantly tighter age constraints. At Galactocentric distances of tens to hundreds of kiloparsecs, and in low-latitude anticenter structures, they typically have apparent magnitudes $r \sim 21$--23. Current 8--10\,m facilities, and even targeted observations with ELT class telescopes, cannot obtain \emph{high-resolution, high signal-to-noise spectra for large samples} of such stars over wide areas of the sky. Without high-quality chemodynamical data for these faint tracers, any reconstruction of the Milky Way's assembly history and halo potential will remain incomplete.

\paragraph{The role of a 30 m-class telescope.}
A 30\,m aperture, combined with high-resolution, highly multiplexed spectroscopy, provides the photon-collecting power and survey efficiency needed to reach these stellar populations. It would enable the acquisition of $\sim 10^{5}$--$10^{6}$ high-quality spectra of faint main-sequence (1--2 mag below the turnoff) and turnoff stars at $r \sim 21$--23 in (i) the northern halo, (ii) numerous stellar streams and shells, and (iii) the outer disk and Galactic anticenter region. Reaching these limits (typically $r\sim 21$ for key tracers) enables precise $v_{\rm rad}$ and MDFs that, together with proper motions, can constrain spatial variations in the Galactic potential and help reconstruct its time-dependent evolution.

\paragraph{The necessity of the northern hemisphere.}
Many of the most informative structures for Galactic archaeology and dark matter studies are located in the \emph{northern} sky: extensive networks of halo streams and shells, a substantial population of ultra-faint dwarf galaxies and low-mass clusters, and the Galactic outer disk and anticenter region, including its warp, flare, and low-latitude overdensities. Robust inference on substructure requires accounting for baryonic perturbers (e.g., giant molecular clouds) and leveraging full chemo-kinematic information.

\paragraph{Representative survey programs enabled by a northern 30 m.}
The capabilities of a northern 30\,m-class facility would allow a set of survey programs that are not feasible with existing or planned instruments:
\begin{itemize}
    \item \textbf{Anticenter and Outer Disk Tomography:} deep spectroscopy of outer disk and disk--halo transition stars to characterize the warp and flare, Monoceros/TriAnd-like structures, and the global halo geometry.
    \item \textbf{Northern Halo Deep Survey:} high-resolution abundances and velocities for $\sim 10^{5}$ faint main-sequence and turnoff halo stars out to $\sim 150$--200\,kpc, decomposing the halo into its constituent accretion events and constraining the total mass and shape of the halo.
    \item \textbf{Northern Streams and Substructure Programme:} chemodynamical mapping of dozens of northern streams to constrain the granularity of the Galactic potential by combining precise $v_{\rm rad}$, MDFs, and proper motions, while accounting for baryonic perturbers.
    \item \textbf{Northern Cluster and Satellite Chronology:} precise ages and detailed abundances for clusters and dwarf galaxies in the outer disk and halo, establishing a time-resolved sequence of the Milky Way's assembly.
\end{itemize}

\paragraph{Synergy with the ELT.}
A northern 30\,m-class facility would provide the essential complement to ESO’s southern ELT, enabling truly all-sky Milky Way archaeology and maximizing synergies with current and future space missions.

\section{Science Goals}

The science case developed in this White Paper can be framed in terms of the following science goals:
\begin{itemize}
    \item \textbf{Milky Way assembly history:} How can we reconstruct, in a quantitatively robust way, the detailed assembly history of the Milky Way by using the chemo-dynamical properties of faint main-sequence (1--2 mag below the turnoff) and turnoff stars in the outer halo and outer disk? Giants are powerful long range tracers, but they are comparatively rare and can yield sparse, selection-biased samples in low surface brightness structures, limiting a fully quantitative reconstruction [7]. Faint main-sequence and turnoff stars provide much higher tracer densities and cleaner chemical tagging, enabling detailed chemodynamical decomposition of the outer halo and disk [9,10].
    \item \textbf{Small-scale structure of dark matter:} What does the fine structure of stellar streams, shells, ultra-faint dwarf galaxies, and outer-disk perturbations---particularly in the halo and Galactic anticenter region---reveal about the granularity of the Galactic potential and the role of substructure when combined with full chemo-kinematic information?
    \item \textbf{A time-resolved, all-sky picture of the Galaxy:} How can we combine measurements for field stars in the MW disk, clusters, and satellites to establish a time resolved, all-sky view of the Milky Way's formation that is fully complementary to, and synergistic with, studies based on southern facilities such as the ELT and space-based surveys?
\end{itemize}

\section{Summary}

By the 2040s–2050s, ongoing and forthcoming astrometric, photometric, and spectroscopic surveys will have revolutionized Milky Way mapping and sharpened the inventory of halo substructure, streams, and satellites. Yet robust constraints on the Galaxy’s detailed assembly and dark matter granularity will remain limited by the lack of high-resolution, highly multiplexed spectroscopy for large samples of faint main-sequence (1--2 mag below the turnoff) and turnoff stars in the anticenter and halo. A northern 30\,m-class telescope equipped with a wide-field, high-resolution, high-multiplex spectroscopic facility would fill this gap and provide the essential complement to southern ELT-class capabilities. It would deliver the radial velocities and multi-element abundances needed to chemodynamically tag early building blocks and to separate accreted debris from perturbed outer disk material in the anticenter. Combined with proper motions, it would map spatial variations in the Galactic potential and enable time-dependent constraints on halo growth through the chemo-kinematic evolution of streams and outer-disk substructure. Beyond survey mapping, the same capability would provide efficient high-resolution follow-up of rare, high-value targets identified by these surveys (e.g., extremely metal-poor stars and the most distant halo tracers), maximizing the scientific return of the 2040s Milky Way data ecosystem.

\section*{References}
{\footnotesize
\setlength{\parindent}{0pt}\noindent
[1] Prusti T. et al., 2016, A\&A, 595, A1,
[2] Hobbs D. et al., 2016, GaiaNIR White Paper, arXiv:1609.07325,
[3] Ivezi\'c {\v Z}. et al., 2019, ApJ, 873, 111,
[4] Laureijs R. et al., 2011, Euclid Definition Study Rep., arXiv:1110.3193,
[5] Spergel D. et al., 2015, WFIRST/AFTA Rep., arXiv:1503.03757,
[6] Gilmozzi R. \& Spyromilio J., 2007, The Messenger, 127, 11,
[7] Freeman K. \& Bland-Hawthorn J., 2002, ARA\&A, 40, 487,
[8] Helmi A. et al., 2018, Nature, 563, 85,
[9] Erkal D. \& Belokurov V., 2015, MNRAS, 450, 1136,
[10] Bonaca A. et al., 2019, ApJ, 880, 38,
[11] Xu Y. et al., 2015, ApJ, 801, 105,
[12] Laporte C.~F.~P. et al., 2018, MNRAS, 481, 286,
[13] Tan C.~Y. et al., 2025, arXiv:2509.12313,
[14] Maunakea Spectroscopic Explorer (MSE) project status update, 2024, MSE/CFHT webpage,
\par}

\end{document}